\RequirePackage{fix-cm}
\documentclass[smallextended]{svjour3}       
\smartqed  
\usepackage{graphicx}
\usepackage{amssymb}
\usepackage{amsmath}
\usepackage{color, colortbl}
\usepackage{mathrsfs}
\usepackage{lscape}
\usepackage{float}
\usepackage{hyperref}

\usepackage{algorithm} 
\usepackage{algpseudocode}

\usepackage{amsmath}
\usepackage{placeins}
\usepackage[inline]{enumitem}
\usepackage[all]{nowidow}

\usepackage{multirow}

\usepackage{url}
\usepackage{setspace}

\usepackage{multicol}

\usepackage[table]{xcolor}

\DeclareMathOperator*{\argmin}{arg\,min}

\definecolor{MAROON}{RGB}{128, 0, 0}
\definecolor{OLIVE}{RGB}{128, 128, 0}
\definecolor{GREEN}{RGB}{0, 128, 0}
\definecolor{NAVY}{RGB}{0, 0, 128}
\definecolor{PURPLE}{RGB}{128, 0, 128}

\usepackage[numbers,sort&compress]{natbib}

\begin{document}

\title{From SIR to SEAIRD: a novel data-driven modeling approach based on the Grey-box System Theory to predict the dynamics of COVID-19
}

\titlerunning {SEAIRD, a novel data-driven modeling approach}       

\author{K. Midzodzi Pékpé   \and
        Djamel Zitouni       \and
        Gilles Gasso         \and
        Wajdi Dhifli         \and
        Benjamin C. Guinhouya*
        }


\institute{Benjamin C. Guinhouya \at
              Univ. Lille, , ILIS, ULR 2694 - METRICS,
              F-59000 Lille, France\\
              Tel.: +33-3-20 623 737\\
              Fax: +33-3-20 623 738\\
              \email{benjamin.guinhouya@univ-lille.fr}           
}

\date{Received: date / Accepted: date}

\maketitle

\begin{abstract}
Common compartmental modeling for COVID-19 is based on a priori knowledge and numerous assumptions. Additionally, they do not systematically incorporate asymptomatic cases. Our study aimed at providing a framework for data-driven approaches, by leveraging the strengths of the grey-box system theory or grey-box identification, known for its robustness in problem solving under partial, incomplete, or uncertain data.
Empirical data on confirmed cases and deaths, extracted from an open source repository were used to develop the SEAIRD compartment model. Adjustments were made to fit current knowledge on the COVID-19 behavior. The model was implemented and solved using an Ordinary Differential Equation solver and an optimization tool. A cross-validation technique was applied, and the coefficient of determination $R^2$ was computed in order to evaluate the goodness-of-fit of the model.
Key epidemiological parameters were finally estimated and we provided the rationale for the construction of SEAIRD model. When applied to Brazil's cases, SEAIRD produced an excellent agreement to the data, with an 
$R^2$ $\geq 90\%$. The probability of COVID-19 transmission was generally high ($\geq 95\%$). On the basis of a 20-day modeling data, the incidence rate of COVID-19 was as low as 3 infected cases per 100,000 exposed persons in Brazil and France. Within the same time frame, the fatality rate of COVID-19 was the highest in France (16.4\%) followed by Brazil (6.9\%), and the lowest in Russia ($\leq 1\%$). 
SEAIRD represents an asset for modeling infectious diseases in their dynamical stable phase, especially for new viruses when pathophysiology knowledge is very limited.

\end{abstract}

\section{Introduction}\label{S:1}
In December 2019, an outbreak of an emerging disease (COVID-19) due to a novel coronavirus, the SARS-CoV-2, began in Wuhan, China and quickly spread in a substantial number of countries \cite{lai2020severe,world2020director}. The COVID-19 pandemic, as a major global health threat, was declared by the WHO on 11 March 2020 \cite{world2020director}. The disease is rapidly spreading in the whole globe, affecting millions of people and pushing governments to take drastic measures to contain the outbreak. For example, mitigation measures to slow transmission through infection prevention and control, and social distancing have been introduced with different timing and pace in countries worldwide. The efficiency of these measures in slowing the transmission of COVID-19 in the general population and, more specifically, in the vulnerable populations of elder adults and individuals with chronic conditions (i.e., hypertension, diabetes, cardiovascular disease, chronic respiratory disease, compromised immune status, cancer and obesity), has been proved useful although the pandemic is still growing. It is noteworthy that once ill of  COVID-19, no treatment with decisive efficiency exists, albeit early supportive therapies can improve outcomes. Thus, preventive strategies and other public health endeavors are to be sustained. One asset to adequately support these interventions may be to get the most clear and realistic picture of the dynamics of the COVID-19 disease, including by taking into account the impact of different mitigation or suppression measures at work in countries. 

Unlike highly data-hungry statistical approaches, which may not be completely suitable in such a situation of data scarcity, common mathematical modeling used in epidemiology for infectious diseases relies on the SIR (Susceptible, Infected, and Recovered or Removed)-type models \cite{kermack1927contribution}, though modeling approaches such as the ARIMA model \cite{Arunkumar2021} coupled with polynomial functions \cite{Hernandez-Matamoros2020}, deep learning \cite{kafieh2021} or even deep learning in combination with compartment model \cite{LiuSEAIRD} have been applied to predict COVID-19 cases. There are many current examples of the application of the compartment modeling in the COVID-19 epidemic \cite{kucharski2020early,wu2020nowcasting,anastassopoulou2020data,roosa2020real}. However, compartment models suffer of a number of issues, including the many a priori assumptions, and the need of a thorough knowledge of the circulating virus, which was difficult at the moment of conducting this study, due to the novelty SARS-CoV-2. In order to compensate the dearth of data and uncertainties around SARS-CoV-2 mechanisms of action and that of its related disease, the COVID-19, we postulate that the grey-box system identification theory (GBSIT) \cite{verhaegen2017, Ljung}, developed in the 1980s 
could make an asset to tackle these challenges. 

Indeed, this theory is one of the most robust ones in situations of prediction and decision-making in the presence of partial, incomplete, or uncertain information \cite{Ljung}. Because of its strong ability to solve uncertain problems, 
and in order to provide good 
predictions under limited knowledge and 
scarce data \cite{yin2013fifteen}, GBSIT appears to be a relevant way to 
describe
the dynamics of 
emerging disease such as COVID-19.
Interestingly, the COVID-19 dynamical model can be seen as a switching system with non-linear modes \cite{verhaegen2017}, where the switches are triggered by control measures set by authorities. As such, the switching model coupled with grey-box approach provides flexibility in characterizing the dynamics of the epidemic across its different phases. Indeed, the grey-box modeling allows to derive the switched system active mode, and must be applied in this case in the epidemic stable dynamic phase (i.e. early stages, before control measures and/or between two measures taken to halt the propagation of a virus) \cite{Ljung, verhaegen2017}. It can also be used in the changing phases of the epidemic (when the mode switches) to estimate trends, and investigate for instance the effectiveness of actions taken to tackle the outbreak. One major advantage of such data-driven approach is its 
ability to operate under limited 
a priori knowledge of the studied phenomenon.

Furthermore, because of their reliance on the calculation of the basic reproduction rate $R_0$ (i.e., the average number of secondary cases arising from a typical primary case in an entirely susceptible population), compartment models often translate into considerable discrepancy between findings \cite{tsai2020american}. $R_0$ is arguably the most important quantity in disease modeling, and there exists a rich mathematical theory supporting how $R_0$ can be computed for a range of SIR-type models with varying degrees of complexity \cite{blackwood2018introduction}.  
However, $R_0$ 
should not be viewed as the ultimate target of modeling so as to be able to estimate any other important parameters (e.g., the rate of infection, recovered people, and deaths in a susceptible population), which often support quick public health responses.

Another concern with compartment models is the difficulty of considering asymptomatic cases. Previous estimates of the proportion of asymptomatic people from COVID-19 provided values between 5\% and 80\% of people being tested positive for COVID-19 but without any symptoms \cite{CEBM2020asymptomatic,mizumoto2020estimating}. However, it is crucial to better characterize the magnitude of the contribution of asymptomatic people to the spread of SARS-CoV-2 in order to be able to develop better strategies to halt this epidemic, not taking into account the possibility of reinfection of some recently infected people \cite{Sharma2020reinfection,Yuan2020reinfection,Carfi2020Persistence,Babikerreinfection}.

 The current study builds upon the work by Hsu and Hsieh, \cite{hsu2008role} who have developed a modeling framework to integrate asymptomatic cases in outbreak dynamics. Our aim is to provide an extension of the SEIR (Susceptible-Exposed-Infected-Removed) compartment model that includes asymptomatic cases in order to better fit both the COVID-19 behavior, as well as the mitigation measures adopted by countries. Unlike usual approach in compartment modeling, another specific feature of the current study is the use of empirical data collected on only the cases and deaths due to COVID-19, – which respectively reflect the transmission and virulence of the virus  – to estimate all the required parameters for the analysis of the COVID-19 dynamics. Finally, the analytical method applied in this study, based on GBSIT, may be sound in supporting policy decisions even if the understanding of the activity of this new virus is only partial, incomplete, or uncertain.


\section{Methods} \label{S:2}
Our modeling approach comprises five components: (i) choice and processing of COVID-19 data; (ii) selection of the most appropriate compartmental modeling for COVID-19; (iii) building the required adaptation to fit current knowledge on the SARS-CoV-2 circulation and transmission; (iv) providing estimation of the targeted parameters using an identification method together with an optimization function; (v) predicting the dynamics of active cases, recovered cases, infective cases, asymptomatic infective cases and the deaths.

\subsection{Setting up and parameterization}
We present a model that considers human transmission of SARS-CoV-2 strain, more unknown than strains of the past with the following assumptions (provided $t$ is the time stamp): (i) Infective persons can be classified in two categories; one of which is with symptoms, denoted by $I(t)$, and the other is without any clinical presence of symptoms called asymptomatic or subclinical infective cases, denoted by $A(t)$. (ii) When a change in the behavior of people occurs due to a public response to the outbreak \cite{hsu2005modeling}, the contact rate (reflecting the level of risky behaviors) decreases with the increase in the cumulative numbers of  removed persons; (iii) Homogeneous mixing population is assumed. Even if our modeling approach is clearly free from key biological assumptions – (a) the birth rate and death rate are equal and given by $\eta$, (b) all individuals are capable of reproducing and are equally subject to mortality, and (c) all individuals are born susceptible to infection –, we keep this in mind in the building and discussion of the modeling strategy.

\begin{figure*}[tp]
\centering\includegraphics[width=0.9\linewidth]{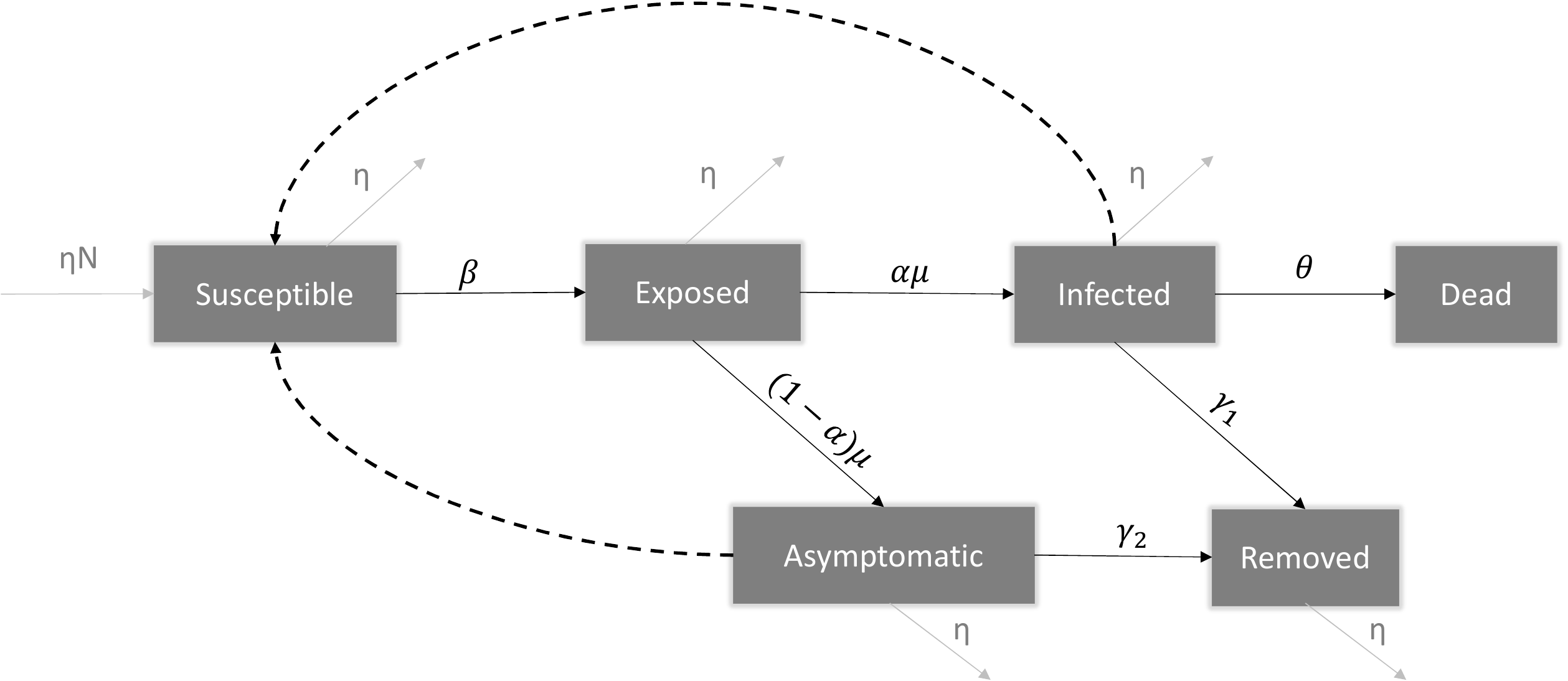}
\caption{Schematic representation of the proposed model.}
\label{fig:fig1}
\end{figure*}

We propose SEAIRD model, a compartment model that explicitly incorporates  asymptomatic cases in the analysis of the epidemic evolution. Our SEAIRD model consists of the following time-dependent variables: $S(t)$: susceptible individuals; $E(t)$: exposed (incubating) population; $A(t)$: asymptomatic infective people; $I(t)$: symptomatic infective individuals; $R(t)$: removed persons (i.e. completely and/or temporary recovered from COVID-19); $D(t)$: Dead people. A flow chart of the model is given in Figure \ref{fig:fig1}. Let us define the parameters as follows:
\begin{itemize}
    \item $\alpha$ is the fraction of exposed population $E$ progressing to class $I$, or proportion of symptomatic infections, with $0<\alpha<1$.
    \item $\beta$ is the probability of COVID-19 transmission (i.e, the infection rate).
    \item $\delta$ is the ratio of the infective force in infectious asymptomatic people by the infective force in infectious (symptomatic) people  
    \item $\gamma_1$ is the recovery probability of infectious people. 
    \item $\gamma_2$ is the recovery probability of asymptomatic people.
    \item $\mu$ is the progression rate from the exposed class to infective class (i.e. infected symptomatic and asymptomatic people).
    \item $\theta$ is the fatality rate.
    \item $\eta$  is the birth and death rate (assumed to be equal here).
    \item N  is the population size of each country.
\end{itemize}

The considered model is a non-linear dynamical model defined as:

\begin{eqnarray*}
\label{eq:our_model}
\frac{dS}{dt} & = &  - \frac{\beta S(t)(I(t) + \delta A(t))}{S(t) + E(t) + I(t) +A(t)+ R(t)}  - \eta S(t) + \eta N \\
\frac{d E}{dt}  & = & \frac{\beta S(t)(I(t) + \delta A(t))}{S(t) + E(t) + I(t) +A(t)+ R(t)}  - (\mu + \eta) E(t) \\
\frac{d I}{dt} & = & \alpha \mu E(t) - (\gamma_1+\theta+\eta) I(t) \\
\frac{d A}{dt} & = & (1 - \alpha) \mu E(t) - (\gamma_2+\eta) A(t) \\
\frac{d R}{dt} & = & \gamma_1 I(t) +\gamma_2 A(t) - \eta R(t) \\
\frac{dD}{dt}&  = & \theta I(t)
\end{eqnarray*}

\noindent For the sake of simplicity, we introduced the so-called state-space vector \cite{verhaegen2017}:
\begin{equation*}
\label{eq:X}
X^\top(t)=\begin{pmatrix}
S(t), & E(t), & I(t), & A(t), & R(t), & D(t)
\end{pmatrix}
\end{equation*}
which gathers the variables at interest.

\subsection{Implementation and optimization}
SEAIRD model involves several parameters to be estimated. We relied on a data-driven approach by leveraging the daily released data to perform the estimation of the parameters. Data on daily infected cases (and probably deaths data) might be subject to noise (i.e. recording errors). Hence, we introduced the cumulative infective cases $\mathscr{I}$ defined as:
\begin{equation*}
\label{eq:def int_I}
\mathscr{I}(t) = \int_0^t{I(\tau)d\tau}.
\end{equation*}
Note that here the integration, which acts as low-pass filter, on variable  $I(t)$ could reduce the noise. 
Similarly, $D(t)$ might be less prone to noise, since $D(t)$ represents the cumulative dead cases on time interval $[0, t]$.
Variables $\mathscr{I}(t)$  and  $D(t)$ are discretized with a sampling period of one day to reduce the computation cost, yielding: 
\begin{equation*}
\label{eq:def int_Id}
\mathscr{I}(t) = \sum_{p=0}^t{I(p)}.
\end{equation*}

Let  $\varphi^\top = 
\begin{pmatrix}
\alpha, &  \beta, & \delta, & \gamma_1, & \gamma_2, & \mu, & \theta
\end{pmatrix}$ be the SEAIRD model vector of parameters.
Based on  empirical data, we can estimate $\varphi$, using the grey-box identification method \cite{verhaegen2017}, an approach widely used for modeling physical systems \cite{Beneventi14}. 
%
Let $y(t)= \begin{pmatrix}
\mathscr{I}(t)  \\ D(t)
\end{pmatrix}$ be the vector including the real cumulative number of infective cases $\mathscr{I}(t) $, and the cumulative number of dead cases $D(t)$ at day $t$. Let $\hat y(t)= \begin{pmatrix}
\hat {\mathscr{I}}(t)  \\ \hat D(t)
\end{pmatrix}$ be the predicted counterpart of $y(t)$ by our model at the same day. 
Since these predictions are obtained with parameters $\varphi$, we write $\hat y(t, \varphi)$. To find the optimal parameters vector which fits our model to the collected data $y(t)$ over a time window $\begin{bmatrix}
T_0, & T_1 \end{bmatrix}$, we considered the following optimization problem over $\varphi$:

$$
\min_{\varphi} \sum_{t=T_0}^{T_1} (w_t^T (y(t) - \hat y(t, \varphi)))^2 
$$
where $w_t \in \mathbb{R}^{2},$ is the weight vector that can be used to balance infected and death cases.

The estimated parameters vector $\varphi$ was finally applied to forecast the course of the COVID-19 disease. The model was implemented and solved using Matlab 2020a, with the Ordinary Differential Equation (ODE45) solver and optimization toolbox. This package implements four different algorithms (Interior Point, Sequential Quadratic Programming, Active Set, and Trust Region Reflective) which serve to estimate SEAIRD model parameters. Two initial conditions were set for this minimization problem: the initial parameters vector $\varphi_{0}$ and the initial state vector $X(0)$. Empirically, we found that $\varphi_{0}$ did not influence the estimation of $\varphi$ contrary to $X(0)$. Here, $X(0)$ was chosen to maximize the coefficient of determination, $0 \leq R^2 \leq 1$. 

\section{Model analysis}
We applied the proposed method (with cross-validation technique \cite{verhaegen2017}) to estimate the COVID-19 dynamics in Brazil. We carried out sensitivity analysis to examine the influence of a likely underestimation of reported number of infected people. Finally, comparative analysis was conducted by estimating similar parameters in other selected  countries.
\subsection{Model estimation and validation}
\begin{figure*}[tp]
\centering\includegraphics[width=1\linewidth]{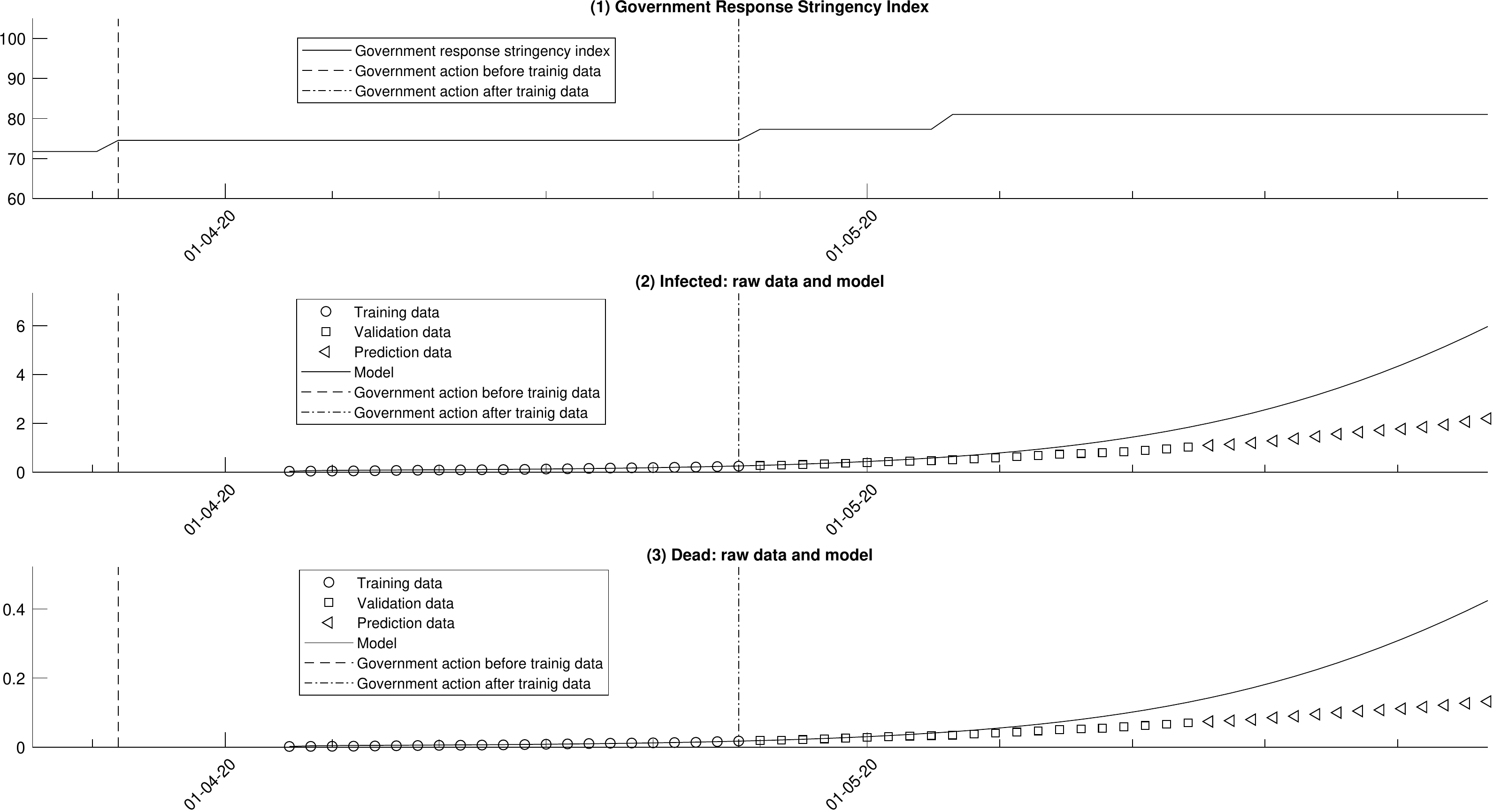}
\caption{Plots of (1) Government Response Stringency Index in Brazil. Time chart on a scale of 0 to 100 (source: \cite{OxCGRT}), (2) cumulative infected for 1000 inhabitants and (3) cumulative deaths for 1000 inhabitants of COVID-19 versus predictions from the model for Brazil.}
\label{fig:brazil_data}
\end{figure*}

The epidemiologic data were taken from an open source repository operated by the European Centre for Disease prevention and Control (ECDC). The database provides daily 
number of new cases and deaths for different countries. 
In order to preclude perturbations to the modeling approach due to changes in behaviors, as a consequence of policy measures taken by countries, we only considered data on the dynamical stable phase of the epidemic, i.e. a period within the setting of control measures (see Figure \ref{fig:brazil_data}). 

The model accuracy was assessed using only data of Brazil, by calculating the coefficient of determination $R^2$ as the goodness-of-fit criterion  of the model. 
As mentioned earlier, we applied the proposed modeling technique to COVID-19 data gathered for Brazil over 42 days from 4 April to 16 May 2020 (see Figure \ref{fig:brazil_data}). This period was chosen because of the relative stability in the control measures taken by the Brazilian government in the response to the COVID-19 outbreak (see Figure \ref{fig:brazil_data}). Thus, the model parameters were estimated in a stable zone as required. Finally, we used 50\% of the collected data for model estimation (4 to 25 April 2020) and 50\% (data from 25 April to 16 May 2020) for the model validation purpose.


The overall modeling procedure is summarized in 
Algorithm \ref{alg:seirad}.
\begin{algorithm}[ht]
	\caption{Parameter estimation of SEAIRD model}
	\label{alg:seirad}
	\textbf{Require}: data $\mathcal{D}=\{y(t)\}_{t=T_0}^{T_f}$ (collected over a period of relative stability in the containment measures, starting at $T_0$ and ending at $T_f$), weights $w$, a threshold $\tau$ where $0 < \tau < 1$ \vspace{3pt} \\
	\textbf{Output}: estimated parameters $\hat \varphi$
	\begin{algorithmic}[1]
		\vspace{6pt}
	    \State Initialize $\varphi_0$
		\vspace{6pt}
		\State Split $\mathcal{D}$ into a training set $\mathcal{D}_{\text{train}}=\{y(t)\}_{t=T_0}^{T_1}$ defined over the time frame $[T_0, T_1]$,  $T_1 < T_f $ and a test set $\mathcal{D}_{\text{test}}=\{y(t)\}_{t=T_1}^{T_f}$ over  $[T_1, T_f]$ \vspace{6pt}
		\State Based on $\mathcal{D}_{\text{train}}$, estimate $\hat{\varphi}$ and the initial state $\hat X(0)$ by solving:
		\begin{equation*}
\hat{\varphi}=\argmin_{\varphi} \left( \min_{X(0)} \sum_{t=T_0}^{T_i} (w_t^T (y(t) - \hat y(t, \varphi)))^2 \right)
\end{equation*}
	\State Validate $\hat{\varphi}$ by computing the fitness measure $R^2$ on $\mathcal{D}_{\text{test}}$ with:
	\begin{equation*}
        R^2 = 1- \frac{\sum_{t=T_1}^{T_f} ( y(t) - \hat y(t, \hat \varphi))^2}{\sum_{t=T_1}^{T_f} ( y(t) - \bar{y}_{test}(t))^2} 
    \end{equation*}
where $\bar{y}_{test}$ is the mean value over $\mathcal{D}_{\text{test}}$ and $\hat y(t, \hat \varphi)$ is the prediction 
    \If{$R^2 \geq \tau$}
		\State Return $\hat{\varphi}$
		\Else 
		\State Go to step 3 and retrain the model starting from a new initialization of $X(0)$
		\EndIf
	\end{algorithmic}
\end{algorithm} 












\color{black}

\subsection{Sensitivity analysis} 
A range of sensitivity analysis was conducted to illustrate the robustness of the model with regards to possible weaknesses linked to data collection on ``confirmed'' cases. In fact, currently reported confirmed cases may be quite far from the actual number of infected cases by COVID-19 \cite{Wu2020confirmedcases}, which might not consider the asymptomatic ones. Therefore, we re-estimated the model parameters and compared its predictive ability by assuming that the actual infected cases might be higher than the reported confirmed cases by: (i) 5\%; (ii) 10\%; and lastly (iii) 20\%. 

\subsection{Estimation for different countries}
The modeling approach developed on the basis of Brazilian data was finally replicated on data gathered for France, India, Russia, South Africa and USA to highlight the common points and differences. Using the World Wide governments response to COVID-19 outbreak time chart \cite{OxCGRT}, the appropriate time window (i.e. the time scale) for the estimation in each country was chosen, so that perturbations to the modeling (due to sudden changes in the control measures decided in each country) can be ruled out. As such, the relevant windows were 26 March 2020 to 16 April 2020, 4 April 2020 to 25 April 2020, 15 April 2020 to 5 May 2020, 4 April 2020 to 25 April 2020, and 4 April 2020 to 25 April 2020 for France, India, Russia, South Africa, and USA, respectively.

\section{Results} \label{S:results}
\subsection{Predictive performance of the model} \label{S:res}
When applied on the data of Brazil, SEAIRD model produced a coefficient of determination $R^2$ of 93\%  and 92\% for the estimation and validation sets, respectively. This suggests a good agreement and adequacy of the model to the data.

\begin{table*}[ht]
    \caption{Parameter estimation from the SEAIRD model. Period of 4 April 2020 to 16 May 2020.} 
    \centering
{\footnotesize
{\rowcolors{4}{lightgray}{white}
    \begin{tabular}{cccccccc}
    \hline
      \multirow{2}{1.3cm}{\centering Parameter}   &  \multirow{2}{1.4cm}{\centering Raw estimate} & \multicolumn{2}{c}{5\% error} & \multicolumn{2}{c}{10\% error} & \multicolumn{2}{c}{20\% error} \\
         \cline{3-8}
         & & \multicolumn{1}{c}{Value} & \multicolumn{1}{c}{Deviation} & \multicolumn{1}{c}{Value} & \multicolumn{1}{c}{Deviation} & \multicolumn{1}{c}{Value} & \multicolumn{1}{c}{Deviation} \\
         \hline
         $\mu$ &  0.0005 & 0.0011 & 129.7\% & 0.0008 & -67.0\% & 0.0009 & 77.9\% \\
         \hline
         $\beta$ &  0.9950 & 0.9951 & 0.01\% & 0.8409 & -15.5\% & 0.9925 & -0.3\% \\
         \hline
  $\delta$ & 49.8 & 40.6& -18.4\% & 40.5 & -17.8\% & 40.7 & -18.23\% \\
  \hline
  $\gamma_1$ &  0.9997 & 0.9949 & -0.5\% & 0.9981 & -0.2\% & 0.9976 & -0.2\% \\
  \hline
$\gamma_2$ & 0.0009 & 0.0035 & 290.2\% & 0.0689 & 7581.2\% & 0.0047 & 423.9\% \\
\hline
$\theta$ &  0.0719  & 0.0691 & -3.9\% & 0.0713 & -0.8\% & 0.0711 & -1.1\% \\
\hline
    \end{tabular}
}
}
    \label{tab:tab1}
\end{table*}

The infection is mainly supported by two major compartments: the infectious and asymptomatic classes. The probability $\beta$ of COVID-19 transmission in Brazil was 99.5\%. In this instance, asymptomatic cases may spread the SARS-CoV-2, 50 times more than symptomatic people. The fatality rate ($\theta$) in Brazil was about 7\% (see Table \ref{tab:tab1}). As shown in Table \ref{tab:tab1}, symptomatic people recovered at a probability of 99\% while the recovery probability of asymptomatic people was 0.09\% (Table \ref{tab:tab1}). 

The sensitivity analysis suggested high deviations to the raw estimates when reported cases of COVID-19 were assumed to be underestimated by 5\%, 10\%, or 20\% (see Table \ref{tab:tab1}). The pattern of the deviation is not clear. In some cases, an underestimation  of the number of reported confirmed cases of COVID-19 of 5\% can produce greater deviations than an error of 10\% or 20\% (e.g., the incidence of infected cases, $\mu$). The effect was the greatest on $\gamma_2$, which can reach up to 7581\% difference when compared to the raw estimates (see Table \ref{tab:tab1}). $\gamma_1$ and $\theta$ were the least affected by potential errors on the confirmed cases. Finally, about the effect of errors on reporting the number of infected cases, $\delta$ is contained in a range of about 20\% below or above.

When compared on the basis of about a 20-day modeling data, the highest differences between countries were found on the $\delta$, $\theta$, $\gamma_2$ parameters (see Table \ref{tab:tab2}). Nonetheless, even if the incidence rate of COVID-19 was generally low across countries, as exemplified by the parameter $\mu$, some differences were apparent. With about 5 cases per 1 million of exposed persons, South Africa exhibited the lowest incidence rate followed by Brazil and France (3 cases per 100,000 exposed persons), Russia (4.5 cases per 100,000 exposed persons), the US (6 cases per 100,000 exposed persons) and India (7 cases per 100,000 exposed persons).
The infective force of asymptomatic cases (as compared to the infective force of symptomatic cases) was the lowest in Russia (5.1) and India (5.3), and the highest in Brazil (42.3), with median values found in South Africa (16.6), France (13.9) and the US (8.5). As shown in Table \ref{tab:tab2}, France exhibited the highest fatality rate from Covid-19 (16.4\%) followed by Brazil (6.9\%). The fatality rate was the lowest in Russia ($\leq 1\%$) (Table\ref{tab:tab2}).

\begin{table*}[tp]
    \centering
    \caption{Estimates from the model in selected countries. Period indicates the training data dates.}
{\footnotesize
{\rowcolors{3}{lightgray}{white}
    \begin{tabular}{ccccccc}
    \hline
         &  Brazil & France & India & Russia & South Africa & United States \\
         \hline
 Period &  4/4 to 25/4 & 26/3 to 16/4 & 4/4 to 25/4 & 15/4 to 5/5 & 4/4 to 25/4 & 4/4 to 25/4 \\
         $\mu$ & 0.0006 & 0.0007 & 0.0016 & 0.0010 & 0.0001 & 0.0014 \\
         \hline
         $\beta$ & 0.7567 & 0.9126 & 0.8840 & 0.9084 & 0.8551 & 0.7492 \\
         \hline
  $\delta$ & 42.3 & 13.9 & 5.3 & 5.1 & 16.6 & 8.5 \\
  \hline
  $\gamma_1$ &   0.9980 & 0.9977 & 0.9962 & 0.9976 & 0.9998 & 0.9993 \\
  \hline
$\gamma_2$ &  0.0682 &0.1002 & 0.1034 & 0.0404 & 0.0001 & 0.0012 \\
\hline
$\theta$ &   0.0698 & 0.1644 & 0.0330 & 0.0097 & 0.0232 & 0.0640 \\
\hline
    \end{tabular}
}
}
    \label{tab:tab2}
\end{table*}

\normalsize

\section{Discussion} \label{S:discu}
We proposed a data-driven approach to estimate key parameters for the COVID-19 epidemic in Brazil and a number of other selected countries. Estimated epidemiological parameters from the model such as the mortality rate or incidence of COVID-19 are consistent with what has been published in the intensive literature on this disease during the last few weeks \cite{CEBM2020globalCFR}. Furthermore, symptomatic people have a high probability to "recover" or to temporary lose the COVID-19 symptoms as reinfection and persistence are newly claimed \cite{Yuan2020reinfection,Carfi2020Persistence}. For asymptomatic individuals, no clear conclusion can be drawn about their recovery potential \cite{Sharma2020reinfection,Babikerreinfection}. Whether, they may continue to support the transmission of the virus with a high load still remain to be explored. Attention should be paid when attempting a direct comparison of estimated parameters between countries. Indeed, the counting practices of COVID-19 cases and testing strategies (e.g., type and number of tests, testing policies), may greatly vary from one country to the other one. Nonetheless, the fatality rates found herein are in line with findings from meta-analytic approach on the topic. The most recent update from the Centre of Evidence-based Medicine (CEBM) of Oxford, as of June, 9, 2020, points out France as the top country in the number of deaths due to COVID-19 at a rate of 18.94\% (95\%CI: 18.75\% - 19.14\%), which is close to the 16.44\% from the SEAIRD model estimation. Equivalent findings for Brazil, India, Russia, South Africa, and the US are 5.25\% (95\%CI: 5.20\% - 5.30\%) and  5.25\% (95\%CI: 5.20\% - 5.30\%), respectively \cite{CEBM2020globalCFR}.

Our analytical strategy was based firstly on the choice of a SIR-type model, which was then adapted to better mirror the behavior of the COVID-19 disease. As a result, we built up the SEAIRD (Susceptible / Exposed /Asymptomatic / Infected / Recovered / Dead) compartments, an extension of the SEIR (Susceptible/Exposed/Infected/Removed) compartment. This adaptation allowed us taking into account: i) the clinically reported latency between the moment of a possible contact with SARS-CoV-2 and the development of COVID-19 symptoms (i.e. the transition from the exposed status to that of an infected symptomatic person), and ii) the importance of asymptomatic (infective) people in the propagation of this virus. Secondly, the mathematical approach used herein was based on the GBSIT \cite{Ljung}. The structure of a grey-box model is built on a combination of knowledge (as white-box models) and empirically collected data (as black-box models). In this context of both liability of knowledge and novelty of the pathogenic agent, grey-box modeling has the potential to take the maximum of advantage of existing data even if they may be partial or incomplete. Thus, with minimal pathophysiological knowledge about the SARS-CoV-2, it was possible to identify important compartments that are then used to determine the transmission pattern and virulence of the COVID-19. Only two variables, empirically collected from April 04, 2020 to May 16, 2020, were needed to derive important parameters that may support public health decision making.

A similar modeling approach, taking into account the compartment of asymptomatic patients, has been recently released by Liu and colleagues \cite{LiuSEAIRD}. The proposal by Liu et al. was unknown to us at submission. 
Nevertheless, as a recall, the modeling approach we propose makes it possible to better portray the changing dynamics of the epidemic according to collective control measures decided by local authorities. The inner principle is that of hybrid systems, which are able to admit different dynamics, depending on the actions they undergo. Secondly, our approach is also based on 
a limited available data (i.e. confirmed cases and deaths) from open access databases in combination with the partial available knowledge at the time of building our SEAIRD model. Finally, as did Liu and coauthors \cite{LiuSEAIRD}, our modeling is not only an endeavor to address the case of asymptomatic people in the spread of COVID-19, but it also has anticipated the possibility of reinfection by the SARS-CoV-2 or some of its various lineages; a projection that was not common in published modeling strategies. Now, the reinfection claim is no longer a hypothesis but a result substantiated by several studies \cite{Yuan2020reinfection,Carfi2020Persistence,Sharma2020reinfection, Babikerreinfection, Dan&Mehtareinfection, Dan1reinfection}.

\begin{figure*}[tp]
\centering\includegraphics[width=1\linewidth]{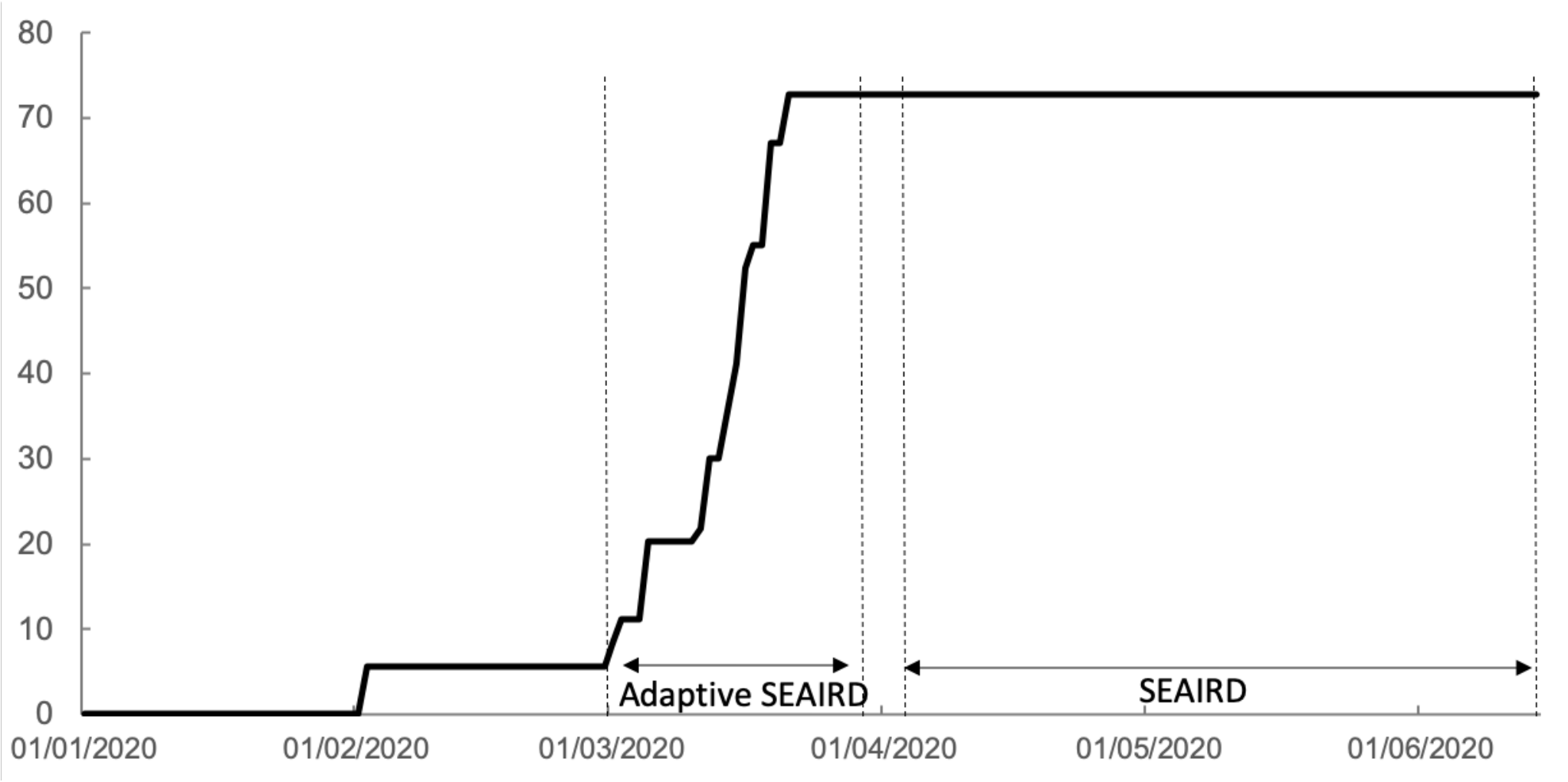}
\caption{Government Response Stringency Index in the US. Periods of changing dynamics phase (``adaptive" SEAIRD model: March, 1, 2020 to March, 29, 2020) vs. stable dynamics phase (The present SEAIRD model: April, 4, 2020 to April, 25, 2020 for modeling followed by accurate predictions up to 2 months later). Time chart on a scale of 0 to 100 (source: \cite{OxCGRT})}
\label{fig:fig3}
\end{figure*}

While the ``adaptive" SEAIRD model of Liu et al. was powered to provide accurate predictions within one to two weeks in advance \cite{LiuSEAIRD}, the predictive ability and scope of the SEAIRD model herein may be longer depending on the duration of the dynamical stable phases of the epidemic. For instance, our SEAIRD model can give accurate prediction for over 2 months in the case of the US (see Supplementary materials). On the other hand, the performance of our SEAIRD model, which is robust for the stable phases, might give poor outputs when used in changing phases of an epidemic dynamics. This standpoint is exemplified when ones attempts to carry out a comparison of results provided by the ``adaptive" SEAIRD \cite{LiuSEAIRD} to those of our SEAIRD model in the period of March, 1, 2020 to March, 29, 2020. As shown in Figure \ref{fig:fig3}, this period falls exactly within a changing dynamics phase of the epidemic in the US, i.e. a time interval separating two control measures decisions. As such, direct comparison with the output of the ``adaptive" SEAIRD model by Liu et al. may be hard, even impossible to perform, due to differences in modeling approach and implementation. Thus, these two different SEAIRD models can be viewed as complementary; one is suitable for the changing phases and the other is for stable phases of an emerging epidemic.

As a hot spot the COVID-19 pandemic at period of conducting the current study, data from Brazil were first put forward for illustrative purpose. Not only is Brazil an emblem of the growth of this pandemic phenomenon \cite{Buss2021Amazon, Candido2020Brazil}, but also because of relative permissiveness and/or flexibility of measures in place, the country appears as a genuine case to learn more about how asymptomatic people are spreading the virus and feeding the pool of ill people. The exact proportion of asymptomatic people is actually unknown due to symptom-based screening that is currently favored. Some studies suggested that between 5\% and 80\% of those who get a positive test for COVID-19 may be asymptomatic \cite{CEBM2020asymptomatic,mizumoto2020estimating}. By definition, an infected person with the disease symptoms may express illness in a way that should not only generate infectious aerosols but also reduce his/her contact with others if sufficiently ill to be in bed. It becomes obvious that the continue transmission of the disease is to be mainly supported by subclinical cases, and especially the asymptomatic people. One consequence of such a situation is that the impact of some public health endeavors, including mitigation or case-patient isolation would be severely diminished \cite{bellgroup2006nonpharmaceutical}.  As in the case of Brazil, these measures are jeopardised by the higher infective force of asymptomatic patients as compared to the symptomatic ones. To gain in efficiency and effectiveness, control measures against COVID-19 should be more stringent, and may stem on the capability of countries in tracing, identifying and caring asymptomatic and/or mild symptoms cases. This is all the most important since recent findings suggested that, contrary to common beliefs, the viral load of SARS-CoV-2 is similar, if not superior, in asymptomatic and mild symptoms patients as compared to their symptomatic counterparts \cite{Zou2020viralload,Argyropoulos2020viralload}. Additionally, earlier findings underscored that up to 55\% of SARS-Cov-2 transmission may be caused by unidentified infected persons \cite{tuite2020estimation, mizumoto2020estimating}. This figure is in line with our results, which show that the infective force of asymptomatic people is about 50 times that of symptomatic people, especially in Brazil.

There are some limitations to our proposed modeling approach. First, it may be less relevant at the very earlier stages/beginning of an epidemic when collected data may be too noisy or too poor to be consistent with the inherent behavior of an epidemic. Furthermore, early data collected in an emerging epidemic such as that of the COVID-19 may not be as good as those collected later due to continuous improvement in field works, as well as in the refinement of diagnostic tools, and so on. Especially, in the case of COVID-19, it should be acknowledged that criteria used to determine the infection status have substantially evolved. From the use of nucleic acid testing, guidance then changed to put emphasis on clinical signs or chest CT scan, and now on serological assays. As a consequence, what is called “confirmed cases” of COVID-19 may somewhat vary according to the type of test(s) used to define them at a specific period of the epidemic. Future studies comparing these different definitions are warranted to secure the comparison of data collected on different time scales. In the meantime, the application of a weighing factor (i.e. forgetting factor) should be pertinent to handle the issue of recent data being more accurate and valid than the earliest ones. Furthermore, as any data-driven approach, the derived parameters from our model depend mainly on the quality of the data used. As such, the fact that our analytical strategy considered only data provided by the European Centre for Disease Control and Prevention (ECDC) might put the outcome of our model under the threat of potential errors in this data repository. A final limitation is associated to the fact that our modeling has assumed a homogeneous population. Such an assumption does not allow taking into account the importance of age groups in the transmission and outcome of this epidemic as compared to findings from recently published study with a different modeling approach on COVID-19 data \cite{Hernandez-Matamoros2020}. A future work is then required, in which new variables (e.g., age pyramid, age-related contamination pattern of the COVID-19) would be considered in the modeling strategy. 

Despite these limitations, our model highly fits data and may describe well the behavior of any  epidemic phenomenon in its dynamical stable phase. Because our SEAIRD model combines simplicity and minimization of the number of input data, which increases its usability and capacity for generalization, we then believe that the proposed approach hold some promises. It can be used not only in the current COVID-19 epidemic, but also generally to future epidemics and notably in the presence of novel viral pathogens for which there may exist neither a treatment nor advanced pathophysiology knowledge. Future development should mandatorily include in the SEAIRD model, the dynamics in different age groups. 

\section*{Declarations}

{\bf Funding sources} \vspace{6pt} \\
This study is not supported by any specific funding.
\vspace{12pt}

\noindent
{\bf Conflicts of interest} \vspace{6pt} \\
We declare no competing interests.

\vspace{12pt}
\noindent
{\bf Ethics approval} \vspace{6pt}\\
Not applicable.

\vspace{12pt}
\noindent
{\bf Consent to participate} \vspace{6pt}\\
Not applicable.

\vspace{12pt}
\noindent
{\bf Consent for publication} \vspace{6pt}\\
Not applicable.

\vspace{12pt}
\noindent
{\bf Availability of data and material} \vspace{6pt}\\
All used epidemiologic data were taken from an \url{https://www.ecdc.europa.eu/en/covid-19-pandemic}{open source repository}  operated by  the  European  Centre  for  Disease  prevention  and  Control  (ECDC).

\vspace{12pt}
\noindent
{\bf Code availability} \vspace{6pt}\\
The executable version of the code is available at \url{https://github.com/midzodzi/Codym/find/main} .

\section*{Acknowledgments}
We would like to thank anonymous teams involved in the data collection and data repositories development all around the world.



%
%

\bibliographystyle{spbasic_unsort}      


\bibliography{sample}

\appendix
\section*{Supplemental material}
\begin{figure}[!h]
\centering\includegraphics[width=1\linewidth]{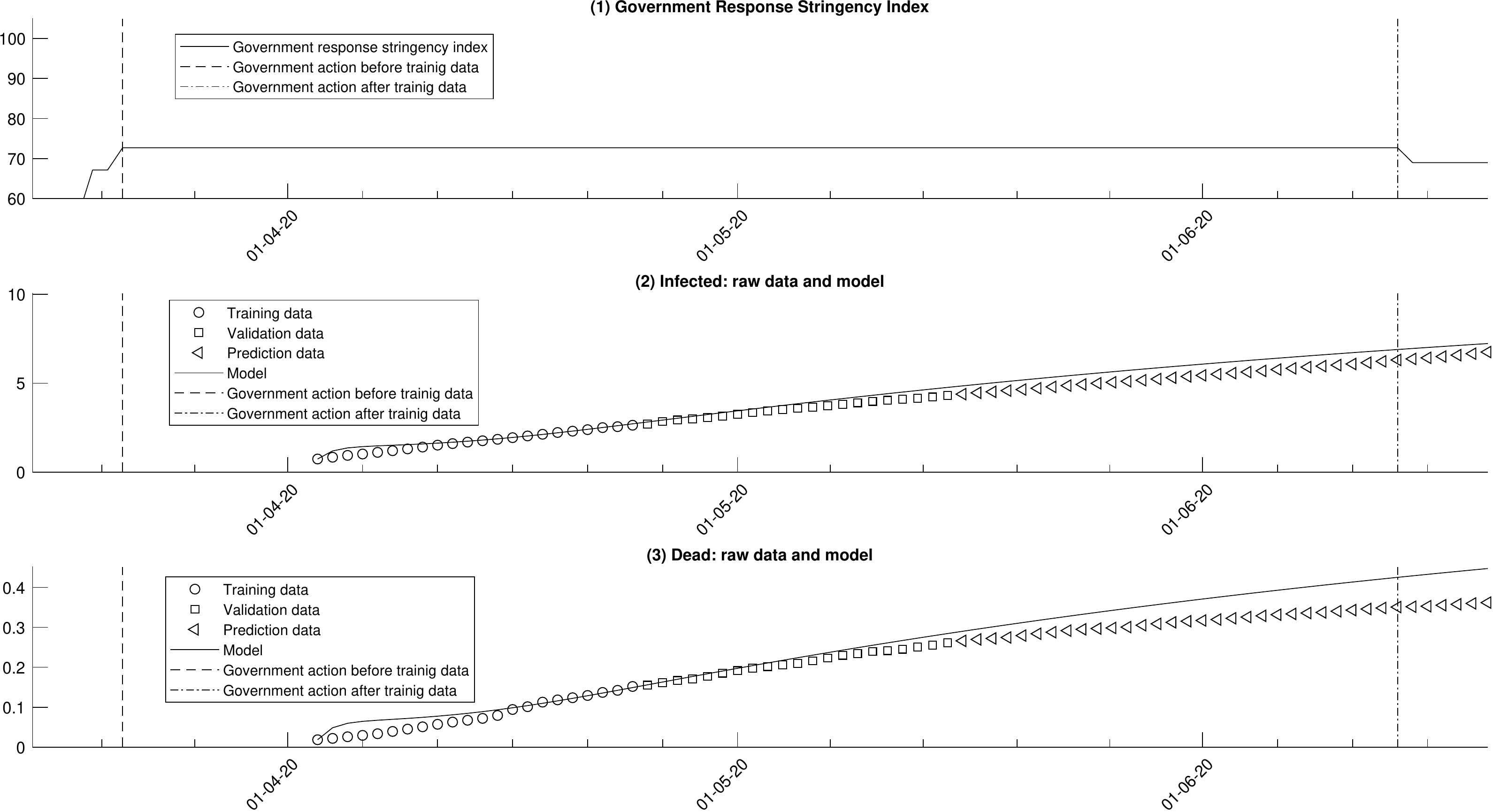}
\caption{Plots of (1) Government Response Stringency Index in the US. Time chart on a scale of 0 to 100 (source: \cite{OxCGRT}), (2) cumulative infected for 1000 inhabitants and (3) cumulative deaths for 1000 inhabitants of COVID-19 vs. predictions from the model for the US.}
\label{fig:US_data}
\end{figure}

\end{document}